\documentclass[
reprint,
amsmath,amssymb,
aps, floatfix,
twocolumn,
pra
]{revtex4-2}

\usepackage{silence}
\usepackage[hidelinks]{hyperref}
\usepackage{comment}
\usepackage{graphicx}
\usepackage{dcolumn}
\usepackage{bm}
\usepackage{float}
\usepackage{xcolor}

\usepackage{amsmath}
\newcommand{\Tr}{\text{tr}}

\begin{document}

\title{Purcell Enhancement and Suppression in Laser Cooling of Yb$^{3+}$:YLF Nanocrystals in a Fabry-Pérot Microcavity
}

\thanks{These authors contributed equally to this work}
\author{Lucas Mendicino $^*$}
 \email{lmendicino@df.uba.ar}
\author{Franco Mayo $^*$}%
 \email{fmayo@df.uba.ar}
\author{Christian Schmiegelow}%
 \email{schmiegelow@df.uba.ar}
\author{Augusto Roncaglia}%
 \email{augusto@df.uba.ar}
\affiliation{%
 Universidad de Buenos Aires, Facultad de Ciencias Exactas y Naturales, Departamento de Física. Buenos Aires, Argentina}%
\affiliation{CONICET - Universidad de Buenos Aires, Instituto de Física de Buenos Aires (IFIBA). Buenos Aires, Argentina}

\begin{abstract}
We investigate the improvement of anti-Stokes laser cooling of a Yb$^{3+}$:YLF nanocrystal in a Fabry-Pérot microcavity via the Purcell effect. Our analysis accounts for both the enhancement of emission lines resonant with the cavity transmission and the suppression of off-resonance emissions. 
Using a quantum-mechanical framework, we modeled the Yb$^{3+}$ ions in a YLiF$_4$ matrix and the laser system to calculate the minimum achievable temperature and cooling efficiency, incorporating cavity-induced modifications to experimental data on emission cross section.
Our results indicate that for temperatures below 100~K, the cooling efficiency ($\eta_c$) is consistently enhanced, and the minimum achievable temperature is reduced comfortably below the current limits. We also show how the inclusion of Purcell inhibition effects can lead to improvements in the cooling efficiency ranging from 25\% to 65\%, with respect to the case when only Purcell enhancement is considered. 
\end{abstract}

\maketitle

\section{Introduction}
\label{sec:intro}

Optical refrigeration of solids has advanced significantly over the past few decades. Since the first demonstrations of laser cooling, this technique has been successfully applied to various systems, the most developed one being rare-earth ion doped crystals~\cite{seletskiy2010laser, seletskiy2011local}. Optical refrigeration has also shown promise for applications such as lowering temperature of quantum devices~\cite{hua2022net} and indirectly cooling devices attached to rare-earth ion doped crystals~\cite{hehlen2018first}. Moreover, optical refrigeration has been observed in levitated nano- and micro-crystals~\cite{laplane2024inert, rahman2017laser, luntz2021laser, ortiz2021laser, zhou2016laser}. The state-of-the-art in optical refrigeration has been achieved using Yb$^{3+}$:YLiF$_4$ crystals, reaching record-low temperatures of 87 K~\cite{volpi2019optical}. Recent proposals have suggested the use of rare-earth dopants, such as holmium (Ho), which have the potential to achieve even lower temperatures~\cite{rostami2019observation}. Additionally, co-doping strategies have been explored~\cite{cittadino2018co}. Although exploring new dopants and crystals is a promising avenue, theoretical evidence shows that modifying the electromagnetic environment of the Yb$^{3+}$:YLiF$_4$, can lead to more efficient cooling and lower temperatures~\cite{ju2024purcell}. Here we improve on these results by considering not only cavity enhancement, but also cavity inhibition. 

\begin{figure}[h!]
    \centering
    \includegraphics[width=0.45\textwidth]{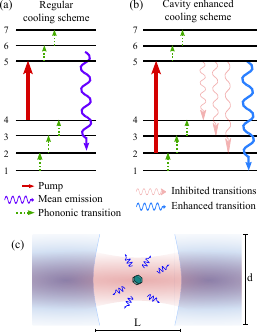}
    \caption{(a) Standard scheme used for anti-Stokes cooling of Yb$^{3+}$:YLF, with pumping on the $4 \leftrightarrow 5$ transition. (b) Scheme of cavity enhanced anti-Stokes cooling, performed by pumping the $2 \leftrightarrow 5$ transition and surrounding the nanoparticle with a Fabry-Pérot micro cavity which enhances the $1 \leftrightarrow 5$ emission line while inhibiting the $\{4,3,2\} \leftrightarrow 5$ transitions. (c) Sketch of a fiber-based Fabry-Pérot microcavity with length $L$ and diameter $d$ where the nanoparticle is levitated at its center.}
    \label{fig:PID}
\end{figure}

In Yb$^{3+}$:YLiF$_4$ crystals, the standard cooling scheme works by pumping from the highest ground-state Stark level to the lowest excited-state Stark level, as illustrated in Fig.~\ref{fig:PID}~(a). As the temperature lowers,  the population of the absorbing level decreases, reducing the cooling efficiency. This limits the minimum achievable temperature, as parasitic absorption begins to dominate over Yb$^{3+}$ ion absorption~\cite{caminati2021loss}. This happens roughly at 80~$\sim$~100~K, when the population of the highest ground state becomes negligible. 

The lifetimes of atomic energy levels can be influenced by the electromagnetic density of states (DOS) in the surrounding environment, a phenomenon known as the Purcell effect~\cite{purcell1946resonance}. The Purcell factor ($F_p$) quantifies the enhancement of an emitter’s spontaneous emission rate relative to its intrinsic value. Importantly, as demonstrated by Heinzen and Feld~\cite{heinzen1987enhanced, hinds1990cavity}, a Fabry-Pérot cavity surrounding an atomic system not only enhances resonant emission but also suppresses off-resonance transitions. High $F_p$ cavities have been utilized to enhance and control emission rates across various systems, including rare-earth ion-doped nanocrystals using bow-tie microcavities~\cite{ju2024purcell}, fiber-based Fabry-Pérot microcavities for NV centers~\cite{kaupp2016purcell,herrmann2023coherent}, solid state  ions~\cite{casabone2021dynamic,deshmukh2023detection, meng2024solid} and semiconductors~\cite{di2012controlling}. 

In this work, we investigate the detailed effects of emission enhancement and suppression in a levitated nanocrystal placed inside a symmetric open fiber Fabry-Pérot microcavity. We consider a nanosized YLF crystal positioned at the node of a concentrical fiber cavity as shown in Fig.~\ref{fig:PID}~(c). The cavity parameters are chosen such that its resonance promotes decay from the excited states only to the lowest of the ground states, enhancing cooling transitions and suppressing others, as exemplified in Fig.~\ref{fig:PID}~(b). The system is pumped from an open side of the cavity at the transition from the second ground state to the lowest excited state.  
With this configuration we examine the enhancement of the emission line in resonance with the cavity as well as the suppression of off-resonance emission lines. Using experimentally measured absorption and emission spectra from Demirbas et al.~\cite{demirbas2021detailed}, which we modify due to the presence of the cavity,
we predict that minimum achievable temperature can be reduced to temperatures below LN$_2$. Moreover, we show that predictions based solely on Purcell effect enhancement underestimate the cooling efficiency if the suppression of off-resonant lines is not considered.

In Section~\ref{sec:model}, we describe the ion-crystal-laser system using a quantum-mechanical formalism and model how the presence of the cavity modifies the full emission spectrum. In Section~\ref{sec:CoolEnh}, we present numerical studies of cooling efficiency for various cavity parameters, comparing them with free-space scenarios.

\section{Model}
\label{sec:model}
\subsection{Laser cooling}

The cooling cycle comprises a combination of three ingredients: optical pumping, the interaction between the ions and the vibrations of the crystal lattice, and radiative relaxation processes. To begin with, let us start by providing a detailed description of the system involved.

We consider the $4$F multiplet of the Yb$^{3+}$ ion which consists of two main manifolds, the ground F$_{7/2}$  manifold and first excited F$_{5/2}$ manifold, as shown in Fig.~\ref{fig:PID}. Transitions between these manifolds have a lifetime in the ms range and can be optically addressed with light in the 950-1050~nm range. Individual levels are Stark-split due to the coupling with the crystal field matrix forming an effective 7-level structure structured with a ground subspace with four levels, and an excited subspace with three levels. As the phononic energies are comparable with the splittings in each sublevel, the interaction with the crystal will cause thermalization within each manifold in ps timescales and slower non-radiative, multi-phonon decays from the excited states to the ground manifold. We will tailor our hamiltonian to account for this dynamics. 

Thus, the hamiltonian for the ion's system is
\begin{equation}
    H_S = \sum_{k=1}^7\,\epsilon_k\, c^\dagger_kc_k,
\end{equation}
where $c^\dagger_k$ and $c_k$ are the fermionic creation and annihilation operators for the $k$th-level. The crystal lattice is described in terms of a phononic reservoir:
\begin{equation}
    H_C = \sum_n\,\hbar\omega_n\,a^\dagger_n a_n, 
\end{equation}
where $a^\dagger_n$ and $a_n$ are bosonic creation and annihilation operators. 

The interaction between the ion and the laser is such that the laser excites it from one of the levels in the ground subspace to its first excited level, i.e. $i\leftrightarrow 5$, with $i\leq 4$. In particular, we will study later the cases $i=4$ and  $i=2$, as illustrated in Fig.~\ref{fig:PID} a) and b). Thus, the hamiltonian describing this process is
\begin{equation}
    H_L = \hbar g\left(c^\dagger_5 c_i + c^\dagger_i c_5\right),
\end{equation}
here $g$ is the Rabi frequency defined as $g=d_0\,E_0/\hbar$, where $d_0$ is the electric dipole moment of the transition $i\leftrightarrow 5$ and $E_0$ the amplitude of the laser field. 

In this way, we can describe the evolution of the ion's internal state in terms of a master equation comprising the interactions with the crystal and the laser~\cite{breuer2002theory, de2018reconciliation}. The crystal is assumed to be much larger than the ion and therefore is always in its equilibrium state at some temperature $T$, and the coupling between crystal and ion is weak in comparison to the energies of the system. Thus, we can obtain the following master equation in Lindblad form for the density matrix $\rho_S$ of the seven-level system:
\begin{eqnarray}
    \dot{\rho}_S &=& -i\left[H_S + H_L, \rho_S\right] + \nonumber \\
    &&+ D_{C}(\rho_S) + D_\mathit{NR}(\rho_S) + D_\mathit{SE}(\rho_S).
\label{ec:master_eq}
\end{eqnarray}
The last three terms correspond to dissipation processes which we can write using the compact form for the Lindblad superoperator $\mathcal{L}(A,\rho) = A\rho A^\dagger - \frac{1}{2}\left\{A^\dagger A, \rho\right\}$.

The first term, $D_C(\rho_S)$, corresponds to the first order transitions caused by the crystal in each of the electronic manifolds and reads:
\begin{eqnarray}    
    D_C(\rho_S) = \sum_{k=1,2,3,5,6} &&\gamma_{\rm ph}(N(\omega_k, \beta)+1)\,\mathcal{L}(c_{k+1}c^\dagger_k,\rho_S)\nonumber +\nonumber \\ &&+ \gamma_{\rm ph}N(\omega_k, \beta)\,  \mathcal{L}(c_k^\dagger c_{k+1}, \rho_S)
\end{eqnarray}
The index runs only over $k=\{1,2,3,5,6\}$ to account only for thermalization within each electronic subspace, $\gamma_{\rm ph}$ describes the phononic relaxation rate and $N(\omega_k,\beta)$ is the thermal distribution for phonons with frequency $\omega_k$ at inverse temperature $\beta=1/k_BT$.

The second term, $D_\mathit{NR}(\rho_S)$, accounts for non-radiative decay  and reads:
\begin{equation}
    D_\mathit{NR}(\rho_S) =\,W_{\rm nr}\, \mathcal{L}(c_5 c^\dagger_4, \rho_S)
\end{equation}
Here we only consider the main non-radiative decay channel, from the electronic level 5 to the level 4 by a multi-phonon transition on the crystal at a rate $W_{\rm nr}$. 

The last term, $D_\mathit{SE}(\rho_S)$, accounts for the spontaneous emission processes and reads:
\begin{equation}
    D_\mathit{SE}(\rho_S) = \sum_{i=5}^7 \sum_{j=1}^4\,\gamma_{i\rightarrow j}\mathcal{L}(c_i c^\dagger_j, \rho_S).
\end{equation}
Here the summation indices reflect the fact that we only consider radiative transitions from the excited F$_{5/2}$ manifold ($i=\{5,6,7\}$) to the ground ones ($j=\{1,2,3,4\}$).

In order to study the thermodynamics of the system we consider the variation of the ion's internal energy:
\begin{align}
    &\dot{E}_S = \Tr{\dot{\rho}_S\,H_S}=\\ \nonumber
    &-\frac{i}{\hbar}\Tr{\left[H_S + H_L, \rho_S\right]H_S}+ \Tr{D_C(\rho_S)\,H_S} \\ \nonumber
    &+ \Tr{D_\mathit{NR}(\rho_S)\,H_S}
    + \Tr{D_\mathit{SE}(\rho_S)\,H_S} .
\end{align}
The first term accounts for the absorbed power, and simplifies to:
$
    P_{\rm abs} = \frac{-i}{\hbar}\Tr{[H_L, \rho_S]H_S},
$
due to the cyclic invariance of the trace. The second and third terms describe the energy current flowing to and from the crystal to the ion; while the second term can be positive or negative and describes the thermalization of each electronic manifold with the crystal, the third term is always negative and describes the energy going from the ion to the crystal via non-radiative decays. Together, they can be used to define the cooling power within the crystal due to the internal dissipation processes as:
$
    P_{\rm cool} = -\Tr{D_C(\rho_S)\,H_S} - \Tr{D_\mathit{NR}(\rho_S)\,H_S}
$.
The last term represents the energy leaving the entire system as emitted photons. Since this energy is always negative, we define the emitted power as: 
$
    P_{\rm emi} =-\Tr{D_\mathit{SE}(\rho_S)\,H_S}.
$
In equilibrium $\dot E_{S}=0$ and can write the following balance equation: 
$
    P_{\rm abs} -P_{\rm emi}=P_{\rm cool}.
$
This reflects the fact that the cooling power is the difference between the energy of the absorbed and emitted photons. 

However, to account for a realistic system, we must take into account that the crystal may contain some impurities, which absorb additional photons and cause a heating effect~\cite{seletskiy2010laser}. Then the relevant quantity we will need to address is the net cooling power defined as: $
    P_{\rm net} = P_{\rm cool} - \alpha_{\rm imp}\,j_0
$
, where $\alpha_{\rm imp}$ is the absorption coefficient of the impurities and $j_0$ is the pump intensity of the laser. Furthermore, we can define the cooling efficiency of the process as the ratio of the net cooling power with respect to the absorbed power as:
\begin{equation}
    \eta=\frac{P_{\rm net}}{P_{\rm abs}}=
    \frac{P_{\rm cool} - \alpha_{\rm imp}\,j_0}{P_{\rm abs}}.
\end{equation} 

To evaluate the properties of the laser cooling process, we examine the steady-state regime where the ion reaches equilibrium.  In this way, we obtain numerically the equilibrium state of Eq.~\eqref{ec:master_eq} for different temperatures of the crystal and intensities of the laser, allowing the characterization of the process for several regimes. 

\subsection{Purcell inhibition and enhancement}
\label{sec:PIE}

To calculate cooling efficiencies and minimum achievable temperatures it is required to supply the model with physical parameters. The spontaneous emission rates $\gamma_{i\rightarrow j}$ in free space were obtained from experimental data~\cite{demirbas2021detailed}. Below, we describe how these values are modified by the presence of a Fabry-Pérot cavity and analyze the improvement in the cooling efficiency.

Introducing the crystal inside a Fabry-Pérot cavity modifies the Ytterbium ions' electromagnetic environment and alters their lifetimes due to the Purcell effect. To account for this, we follow the approach described by Feld et. al.~\cite{heinzen1987enhanced}. The presence of the cavity modifies the spontaneous emission rates as:
\begin{equation}
    \frac{\gamma}{\gamma_{\rm sp}} = \frac{1}{1 - R} \, \frac{1}{1 + [1 / (1-R)]^2 \sin^2{(kL)}},
\end{equation}
where $\gamma$ is the spontaneous-emission rate into the cavity, $\gamma_{\rm sp}$ the free-space rate into the same solid angle, $R$ represents the mirrors' reflectivity (which we consider equal for both mirrors), $k$ is the wave-number of the pumping light, and~$L$ the cavity length. This expression has been obtained through a 1-dimensional approximation, considering cavity mirrors with finite length and that the total intensity of the Purcell effect depends on how much of the environment is modified by the cavity. Thus, the total emission rate $\Gamma$ is:

\begin{equation}
    \frac{\Gamma}{\Gamma_{\rm sp}} = 1 + \left(\frac{\gamma}{\gamma_{\rm sp}} - 1 \right) \frac{3}{8 \pi} \Delta \Omega,
    \label{ec:Gamma_total}
\end{equation}
where $8 \pi / 3$ is the angular integral of the radiation of a dipole emitter~\cite{jackson_classical_1999} and $\Delta \Omega$ is the solid angle subtended by the cavity mirrors around the nanoparticle~\cite{heinzen1987enhanced}, which can be expressed in terms of the cavity length $L$ and diameter $d$ as $\Delta \Omega = 8 \pi \cdot (d/ 2 L)^2$.

\begin{figure}[b!]
    \centering
    \includegraphics{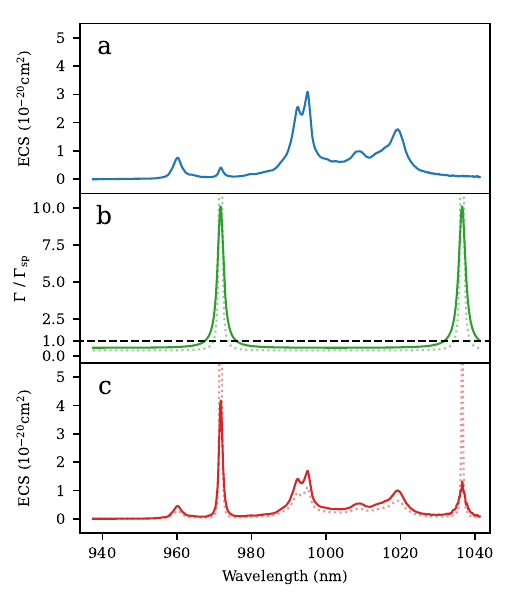}
    \caption{
    (a) Emission cross section $\Gamma_\text{sp}$ spectra of a Yb$^{3+}$:YLF crystal at~150~$K$, obtained from Demirbas et. al.~\cite{demirbas2021detailed}. (b) Calculated relative emission rate $\Gamma/\Gamma_\text{sp}$  for a cavity of length $L=$~$16$~$\cdot$~$\lambda_{1 \leftrightarrow 5}/2$ in resonance with the $1 \leftrightarrow 5$ emission line, with a mirror diameter and reflectivity of $d$~=~$3.11$~$\mu$m and $R$~=~$0.95$ (solid), and  
    $d$~=~$3.61$~$\mu$m and $R$~=~$0.99$ (dashed).
    (c) Modified emission cross section $\Gamma$ of a Yb$^{3+}$:YLF nanocrystal in the presence of the cavity as in (b).}
    \label{fig:spectra_purcell}
\end{figure}

In Fig.~\ref{fig:spectra_purcell} we calculate the Purcell modified emission spectra as a convolution between the emission of the Yb$^{3+}$:YLF measured in~\cite{demirbas2021detailed} and the total emission rate of two different FP cavities as described in Eq.~\eqref{ec:Gamma_total}. We consider cavities with the same length~$L$~$=$~$7.7736$~$\mu$m~$=$~$16$~$\cdot$~$\lambda_{1 \leftrightarrow 5}/2$, tuned at the $1 \leftrightarrow 5$ transition, but different diameters, one with~$d$~$=$~$3.11$~$\mu$m (continuous line) and the other with~$d$~$=$~$3.61$~$\mu$m (dotted line).

We aim to modify the emission rates using a Fabry-Pérot cavity to inhibit the least energetic radiative decays and enhance the most energetic one at $\sim$971~nm, as depicted in Fig.~\ref{fig:PID}. For this purpose, we choose a cavity with an appropriate free spectral range of least~$\sim$~$60$~nm. This ensures that unwanted transitions are not enhanced, which could lead to undesirable heating of the crystal, particularly when its emission wavelength lies between~$\sim$~$977.7$~nm and~$\sim$~$1030$~nm. Consequently, this imposes an upper limit on the cavity length~$L$ of approximately~$L$~$\sim$~$7.77\;\mu$m.

In figure Fig.~\ref{fig:spectra_purcell}, we show the emission of the Yb$^{3+}$:YLF measured in~\cite{demirbas2021detailed}, the cavity relative emission rate, and the convolution of these two as a function of the wavelength for a cavity of length~$L=$~$16$~$\cdot$~$\lambda_{1 \leftrightarrow 5}/2$. We can observe how it effectively inhibits all unwanted transitions while it enhances the desired one. In the following section, we discuss how the use of this modified spectrum changes the cooling efficiencies and minimum achievable temperatures. 

\section{Cooling enhancement}
\label{sec:CoolEnh}

\begin{figure}[b!]
    \centering
    \includegraphics{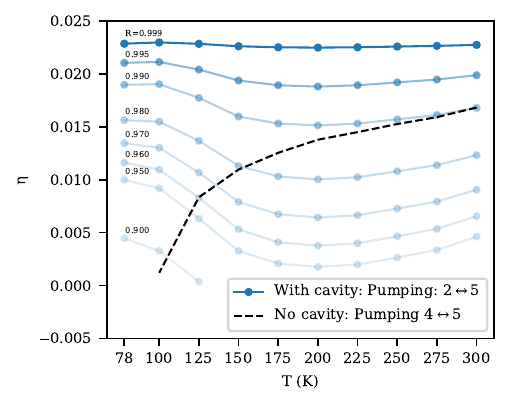}
    \caption{Efficiency at maximum cooling power as a function of temperature. In blue, the cavity enhanced scheme with pumping on the $2\leftrightarrow 5$ transition, for different mirror reflectivities and for a cavity with L=~$16$~$\cdot$~$\lambda_{1 \leftrightarrow 5}/2$ and $d$~=~$3.11$~$\mu$m. In black the standard cooling scheme without cavity, pumping in the $4\leftrightarrow 5$ transition.}
    \label{fig:eff_vs_T}
\end{figure}

In order to study the laser cooling process, we use the following parameters: $\gamma_{\rm ph}=10^{12}\;{\rm s^{-1}}$, $d_0=6.6\times10^{-31}\;{\rm C\,cm}$, $w_{\rm nr}=1.45\; \rm s^{-1}$ and $\alpha_{\rm imp}=4\times10^{-4}\;{\rm cm^{-1}}$ along with the spectral information from~\cite{demirbas2021detailed}. These parameters are similar to those used in~\cite{ivanov2016minimization} and allow us to recover the results of the semiclassical model in the absence of a cavity~\cite{seletskiy2008cooling}. Regarding the cavity, we consider a length~$L=$~$16$~$\cdot$~$\lambda_{1 \leftrightarrow 5}/2$, a diameter $d=3.11\;\mu $m, and several values for the reflectivity of the mirrors.

\begin{figure*}[t!]
    \centering \includegraphics{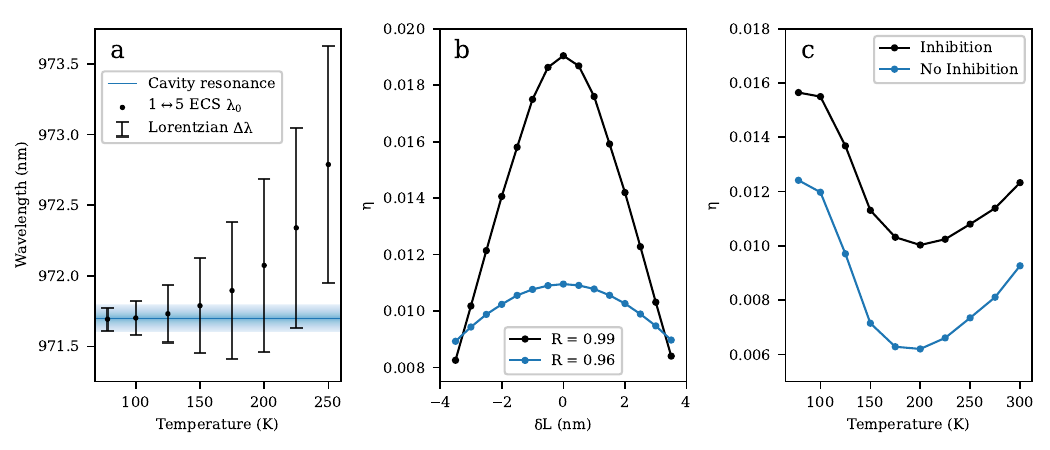}
    \caption{(a) Wavelength matching of the $1 \leftrightarrow 5$  emission line to the cavity as a function of temperature. 
    In black, the central emission wavelength and Lorentzian width from Demirbas et. al.~\cite{demirbas2021detailed}.
    In blue shade, the transmission spectrum of a cavity designed to be in resonance to the $1 \leftrightarrow 5$ at low temperatures, with $R= 0.99$.
    (b) Cooling efficiency at $T=100 K$ as a function of the cavity length mismatch $\delta L$ expressed in {nm} for $R=0.96$ and $R=0.99$. Efficiency drops as the cavity becomes detuned from the $1\leftrightarrow 5$ transition.
    (c) Efficiency at maximum cooling power vs temperatures for a cavity with reflectivity $R=0.98$ when considering (blue) and not considering (black) inhibition. We observe that the efficiency is always higher when considering inhibition by amounts ranging from $25\%$ to $65\%$.
    For all three plots the cavity diameter and length were $d$~=~$3.11$~$\mu$m,~$L=$~$16$~$\cdot$~$\lambda_{1 \leftrightarrow 5}/2$.
    }
    \label{fig:triad}
\end{figure*}

The impact of the cavity on the cooling efficiency $\eta$ can be seen in Fig.~\ref{fig:eff_vs_T}. There we plot the cooling efficiency as a function of temperature, considering cavities made with mirrors of varying reflectivity and pumping the $2\leftrightarrow 5$ transition. We can see that the presence of the cavity not only enhances the cooling efficiency in the low temperature range, but also broadens the temperature limit over which cooling is effective. 
As expected, the efficiency increases with the reflectivity of the mirrors. Remarkably, a significant enhancement can still be observed for experimentally achievable reflectivities up to $0.99$.
We find that cooling persists at temperatures well below 100~K, which is the temperature at which the cooling ceases in the absence of the cavity.  
Our results are restricted to temperatures above 78~K due to a lack of data on the crystal spectrum at lower temperatures. However, the tendency clearly indicates that the minimum achievable temperature will be significantly reduced by the use of the cavity.

In Fig.~\ref{fig:eff_vs_T}, we also show the cooling efficiency curve without the cavity, which is close to zero at a temperature of 100~K. The plotted result corresponds to a pump tuned to the $4\leftrightarrow 5$ transition, which is the best cooling strategy without a cavity. 
We evaluated the model with the laser tuned either to the $2\leftrightarrow 5$ or $3\leftrightarrow 5$ transitions and verified that it raises the minimum achievable temperatures, consistent with previous results~\cite{seletskiy2010laser}. Moreover, we also analyzed the case of two pumping lasers on different transitions and concluded that, without a cavity, the best cooling cycle is achieved with one laser tuned to the $4\leftrightarrow 5$ transition.

In the absence of a surrounding cavity, the efficiency exhibits a characteristic monotonic behavior. However, as shown in Fig.~\ref{fig:eff_vs_T}, the presence of the cavity leads to a marked increase in efficiency near 78~K. This enhancement arises because the cavity was specifically designed to be optimally resonant with the $1 \leftrightarrow 5$ transition at low temperatures. Due to the temperature-dependent spectral shift and broadening of the transition, optimal cooling is achieved only when the emission line aligns with the cavity resonance in both wavelength and linewidth. This effect is illustrated in Fig.~\ref{fig:triad}~(a), where we plot the central wavelength and linewidth of the $1 \leftrightarrow 5$ emission as a function of temperature and compare them with the cavity spectrum. As the temperature decreases, the emission gradually becomes resonant with the cavity, resulting in enhanced cooling efficiency.

Naturally, the length of the cavity plays a critical role in determining the cooling efficiency, as it directly affects the spectral alignment between the cavity modes and the relevant transitions. In Fig.~\ref{fig:triad}~(b), we show how the efficiency varies with cavity length for two different mirror reflectivities. These results highlight the importance of precise length control: small deviations from the optimal value can significantly reduce efficiency, with higher reflectivity leading to greater sensitivity to detuning.

Finally, to illustrate the overall effect of cavity-induced modifications on the emission spectrum, Fig.~\ref{fig:triad}~(c) shows that neglecting inhibition leads to a significant reduction in cooling efficiency. This is because transitions that contribute to heating are no longer suppressed by the cavity. We emphasize that it is essential to account not only for the enhancement provided by the Purcell effect, but also for the inhibitory effects it introduces. Calculations that ignore inhibition yield efficiencies that are 25\% to 65\% lower, depending on the temperature, than those obtained when inhibition is properly included.

\section{Conclusions}

In this work, we have shown that the Purcell effect can be harnessed to enhance both the cooling efficiency and the minimum achievable temperature of Yb$^{3+}$:YLF nanocrystals embedded in a Fabry-Pérot microcavity. Our analysis reveals that modifications by the cavity in the emission spectrum enhance cooling at low temperatures compared to free-space conditions. Importantly, we emphasize that this improvement arises not only from the enhancement of desired transitions but also from the inhibition of unwanted, heating transitions. Neglecting these inhibitory effects can lead to an underestimation of the efficiency by up to 65\%, depending on the temperature.

While our results suggest a lower minimum achievable temperature in the cavity-assisted regime, a precise prediction requires emission data at temperatures below 78~K. Nevertheless, the method is expected to reduce the minimum temperature by at least a factor of two, since the lower-energy level involved in the $1 \leftrightarrow 5$ transition (as opposed to the $1 \leftrightarrow 4$ in free space) is thermally depopulated at significantly lower temperatures.

Finally, we find that the optical cavity should have a length of at least $L$~$\sim$~$7.77$~$\mu$m, corresponding to a free spectral range of about 60~nm, sufficient to isolate a single emission line while suppressing others. Such cavities are compatible with existing fiber-based microcavity technology, making this approach experimentally feasible.

\begin{acknowledgments}

This work was supported by Agencia I+D+i Grants No. PICT 2018 - 3350, No. PICT 2019 - 4349 and PICT-2021-01288, Secretaría de Ciencia y Técnica, Universidad de Buenos Aires Grant No. UBACyT 2018 (20020170100616BA) and No. UBACyT (20020130100406BA), and CONICET (Argentina).

We would like to acknowledge Umit Demirbas and his group for generously sharing their data with us~\cite{demirbas2021detailed}.
\end{acknowledgments}

\bibliography{ref}

\end{document}